\documentclass[usenatbib,twocolumn]{aastex631}
\usepackage{xurl}
\usepackage{subfigure}
\usepackage{silence}
\usepackage{hyperref}
\usepackage{graphicx}	
\usepackage{amsmath}	
\usepackage{xspace}	
\usepackage{color}
\usepackage[utf8]{inputenc}
\newcommand{\sw}[1]{\texttt{#1}}

\newcommand{\elec}{e^{-}}
\shorttitle{GROWTH-India Telescope}
\shortauthors{Harsh Kumar et al.}

\begin{document}
\title{India's first robotic eye for time domain astrophysics: the GROWTH-India telescope}
\author[0000-0003-0871-4641]{Harsh Kumar}
\email{harshkumar@iitb.ac.in}
\affiliation{Department of Physics, Indian Institute of Technology Bombay, Powai, 400 076, India}

\author[0000-0002-6112-7609]{Varun Bhalerao}
\affiliation{Department of Physics, Indian Institute of Technology Bombay, Powai, 400 076, India}

\author[0000-0003-3533-7183]{G.C. Anupama}
\affiliation{Indian Institute of Astrophysics, 2nd Block 100 Feet Rd, Koramangala Bangalore, 560 034, India}

\author[0000-0002-3927-5402]{Sudhanshu Barway}
\affiliation{Indian Institute of Astrophysics, 2nd Block 100 Feet Rd, Koramangala Bangalore, 560 034, India}

\author[0000-0001-7570-545X]{Judhajeet Basu}
\affiliation{Indian Institute of Astrophysics, 2nd Block 100 Feet Rd, Koramangala Bangalore, 560 034, India}

\author[0000-0001-5253-3480]{Kunal Deshmukh}
\affiliation{Department of Metallurgical Engineering and Materials Science, Indian Institute of Technology Bombay, Powai, Mumbai-400076, India}
\affiliation{Department of Physics and Astronomy, Texas Tech University, PO Box 41051, Lubbock TX 79409, USA}

\author[0000-0002-8989-0542]{Kishalay De}
\affiliation{Division of Physics, Mathematics and Astronomy, California Institute of Technology, Pasadena, CA 91125, USA}
\affiliation{MIT-Kavli Institute for Astrophysics and Space Research, 77 Massachusetts Ave., Cambridge, MA 02139, USA}

\author[0000-0002-7708-3831]{Anirban Dutta}
\affiliation{Indian Institute of Astrophysics, 2nd Block 100 Feet Rd, Koramangala Bangalore, 560 034, India}
\affiliation{Pondicherry University, R.V. Nagar, Kalapet, 605014, Puducherry, India}

\author[0000-0002-4223-103X]{Christoffer Fremling}
\affiliation{Division of Physics, Mathematics and Astronomy, California Institute of Technology, Pasadena, CA 91125, USA}

\author[0000-0002-2395-8727 ]{Hrishikesh Iyer}
\affiliation{Department of Electrical and Computer Engineering, University of Illinois at Urbana-Champaign, Urbana, IL 61801, USA}

\author[0000-0003-2191-4541]{Adeem Jassani}
\affiliation{Department of Physics, Indian Institute of Technology Bombay, Powai, 400 076, India}

\author[0000-0002-8875-6413]{Simran Joharle}
\affiliation{Fergusson College, Shivajinagar, Pune, Maharashtra, 411 004, India}
\affiliation{Heidelberg University, Grabengasse 1, 691 17, Heidelberg, Germany}

\author[0000-0003-2758-159X]{Viraj Karambelkar}
\affiliation{Department of Physics, Indian Institute of Technology Bombay, Powai, 400 076, India}
\affiliation{Division of Physics, Mathematics and Astronomy, California Institute of Technology, Pasadena, CA 91125, USA}

\author{Maitreya Khandagale}
\affiliation{Department of Physics, Indian Institute of Technology Bombay, Powai, 400 076, India}
 
\author{K Adithya Krishna}
\affiliation{Department of Physics, Indian Institute of Technology Bombay, Powai, 400 076, India}

\author[0000-0001-8057-0203]{Sumeet Kulkarni}
\affiliation{204 Lewis Hall, The University of Mississippi, Oxford MS 38677, United States of America}

\author[0000-0001-5536-4635]{Sujay Mate}
\affiliation{Department of Astronomy and Astrophysics, Tata Institute of Fundamental Research, Mumbai, 400005, India}

\author{Atharva Patil}
\affiliation{Graduate Institute of Astronomy, National Central University, 32001, Taiwan}

\author[0000-0002-1762-0834]{DVS Phanindra}
\affiliation{Indian Institute of Astrophysics, 2nd Block 100 Feet Rd, Koramangala Bangalore, 560 034, India}

\author{Subham Samantaray}
\affiliation{Department of Physics, Indian Institute of Technology Bombay, Powai, 400 076, India}

\author[0000-0002-4477-3625]{Kritti Sharma}
\affiliation{Department of Mechanical Engineering, Indian Institute of Technology Bombay, Powai, Mumbai-400076, India}

\author[0000-0003-4531-1745]{Yashvi Sharma}
\affiliation{Department of Physics, Indian Institute of Technology Bombay, Powai, 400 076, India}
\affiliation{Division of Physics, Mathematics and Astronomy, California Institute of Technology, Pasadena, CA 91125, USA}

\author{Vedant Shenoy}
\affiliation{Department of Physics, Indian Institute of Technology Bombay, Powai, 400 076, India}

\author[0000-0003-2091-622X]{Avinash Singh}
\affiliation{Indian Institute of Astrophysics, 2nd Block 100 Feet Rd, Koramangala Bangalore, 560 034, India}
\affiliation{Hiroshima Astrophysical Science Center, Hiroshima University, Higashi-Hiroshima, Hiroshima, Japan - 739-8526}

\author[0000-0003-4524-6883]{Shubham Srivastava}
\affiliation{ Astrophysics Research Centre, School of Mathematics and Physics, Queen's University Belfast, Belfast, BT7 1NN, UK}

\author[0000-0002-7942-8477]{Vishwajeet Swain}
\affiliation{Department of Physics, Indian Institute of Technology Bombay, Powai, 400 076, India}

\author[0000-0003-3630-9440]{Gaurav Waratkar}
\affiliation{Department of Physics, Indian Institute of Technology Bombay, Powai, 400 076, India}

\author{Dorje Angchuk}
\affiliation{Indian Astronomical Observatory, Indian Institute of Astrophysics, Post Box No. 100, Leh-Ladakh (UT) 194 101 India}

\author{Padma Dorjay}
\affiliation{Indian Astronomical Observatory, Indian Institute of Astrophysics, Post Box No. 100, Leh-Ladakh (UT) 194 101 India}

\author{Tsewang Dorjai}
\affiliation{Indian Astronomical Observatory, Indian Institute of Astrophysics, Post Box No. 100, Leh-Ladakh (UT) 194 101 India}

\author{Tsewang Gyalson}
\affiliation{Indian Astronomical Observatory, Indian Institute of Astrophysics, Post Box No. 100, Leh-Ladakh (UT) 194 101 India}

\author{Sonam Jorphail}
\affiliation{Indian Astronomical Observatory, Indian Institute of Astrophysics, Post Box No. 100, Leh-Ladakh (UT) 194 101 India}

\author{Tashi Thsering Mahay}
\affiliation{Indian Astronomical Observatory, Indian Institute of Astrophysics, Post Box No. 100, Leh-Ladakh (UT) 194 101 India}

\author{Rigzin Norbu}
\affiliation{Indian Astronomical Observatory, Indian Institute of Astrophysics, Post Box No. 100, Leh-Ladakh (UT) 194 101 India}

\author[0000-0002-2966-2951]{Tarun Kumar Sharma}
\affiliation{Physikalisches Institut der Universität zu Köln, Zülpicher Strasse 
77, 50937 Cologne, Germany}

\author{Jigmet Stanzin}
\affiliation{Indian Astronomical Observatory, Indian Institute of Astrophysics, Post Box No. 100, Leh-Ladakh (UT) 194 101 India}

\author[0000-0002-1469-3958]{Tsewang Stanzin}
\affiliation{Indian Astronomical Observatory, Indian Institute of Astrophysics, Post Box No. 100, Leh-Ladakh (UT) 194 101 India}

\author{Urgain Stanzin}
\affiliation{Indian Astronomical Observatory, Indian Institute of Astrophysics, Post Box No. 100, Leh-Ladakh (UT) 194 101 India}

\begin{abstract}
We present the design and performance of the GROWTH-India telescope, a 0.7~m robotic telescope dedicated to time domain astronomy. The telescope is equipped with a 4k back-illuminated camera gives a 0.82\degr\ field of view and a sensitivity of $m_\mathrm{g^\prime}\sim$20.5 in 5-min exposures. Custom software handle observatory operations: attaining high on-sky observing efficiencies ($\gtrsim 80\%$) and allowing rapid response to targets of opportunity. The data processing pipelines are capable of performing PSF photometry as well as image subtraction for transient searches. We also present an overview of the GROWTH-India telescope's contributions to the studies of Gamma-ray Bursts, the electromagnetic counterparts to gravitational wave sources, supernovae, novae and solar system objects.
\end{abstract}

\section{Introduction}
Time-domain astrophysics (TDA) has become a major thrust area around the world, and is poised to continue growing in the coming decade \citep{decadal04,de2007science,nwnh10,2021pdaa.book.....N}. The studies of transient sources allow us to probe some of the most extreme environments in the universe and, in turn, address open questions from such diverse fields as the distributions of planets around other stars, the fates of massive stars, the nature of accretion and jets, and nucleosynthesis in the universe \citep[see for instance][and references therein]{laa+09,kasliwal2011}. Time-domain studies span many orders of magnitudes in timescales. In optical bands, new high speed instruments study variability at sub-second timescales \citep[etc.]{2007MNRAS.378..825D,2016SPIE.9908E..0YD}, while long-term variability studies have spanned decades \citep{2009ASPC..410..101G,2015MNRAS.453.1562G}.

The varied classes of targets studied in TDA each have their specific requirements from observational facilities. Many sources like supernovae, novae, period objects like binaries, etc., require monitoring at a cadence of one or a few days: a task that is cumbersome for classically scheduled telescopes but straightforward for queue-scheduled ones. Fast transients like afterglows of Gamma-Ray Bursts (GRBs) or electromagnetic counterparts to gravitational wave sources (EMGW) demand fast responses within minutes or even seconds of discovery of the transient. They also require a geographically distributed network of observatories to ensure that the target is visible to at least some of them. Some sources require continuous monitoring, unhindered by the diurnal cycle. This too, requires a geographically spread out set of observatories coordinating their efforts. To address these issues, various global networks have been created with telescopes geographically distributed around the world --- for instance, the Global Relay of Observatories Watching Transients Happen \citep[GROWTH;][]{2019PASP..131c8003K}, Las Cumbres Observatory Global Telescope \citep[LCOGT;][]{2013PASP..125.1031B}, Global Rapid Advanced Network Devoted to the Multi-messenger Addicts \citep[GRANDMA;][]{2020MNRAS.492.3904A}, and MASTER \citep{2010AdAst2010E..30L}. Typically, such networks have had more telescopes in the ``western hemisphere'', with limited participation from ``eastern'' countries.
Indian astronomers have been active in time-domain astrophysics using a variety of optical and infra-red facilities in the country (for a brief overview of the facilities, see \citealt{2017PINSA..87....1S} and references therein).
However, these facilities are classically scheduled and heavily subscribed, with time-domain observations largely executed through a limited number of Target-of-Opportunity (ToO) triggers. 

This motivated us to set up the GROWTH-India Telescope (GIT), a fully autonomous dedicated telescope for time-domain astrophysics. The facility was set up with certain key goals in mind. The main focus was to have a fully automated system that can respond to fast transient triggers with minimal lag, including autonomous responses to GCN \citep{Barthelmy2003} and similar triggers. A closely aligned goal was to create automated and standardised pipelines for all data processing so that the team can focus on interpretation rather than data reduction. The telescope needed to be sensitive to about 21~mag in optical, with $\sim 1^{\circ}$ field of view. We opted for an off-shelf telescope and camera for rapid deployment, high reliability, and ready availability of technical support. GIT would work closely with partners in the

GROWTH sharing the key science goals (\S\ref{sec:science}). 

In this paper, we describe the configuration, performance, and some scientific results from GIT. The paper is organised as follows. In \S\ref{sec:config} we describe the site, telescope, and the instruments. \S\ref{sec:obs} deals with the nightly autonomous observing procedures, while the data processing is described in \S\ref{sec:dataproc}. The sensitivity and performance of the telescope are discussed in \S\ref{sec:perf}. Lastly, \S\ref{sec:science} enumerates the key science goals of GIT and presents some results from the first few years of operation.

\section{System configuration}\label{sec:config}

GIT is a 70-cm telescope designed for autonomous operations, backed up by local and remote observing capabilities. The primary camera is an Andor iKon-XL 230 CCD camera with a 16.8-megapixel back-illuminated sensor.
The telescope and camera have been chosen to provide a unique combination of a $0.7\degr$ wide field of view (FoV) and good sensitivity. The telescope was installed and commissioned in the summer of 2018 and is jointly operated by the Indian Institute of Astrophysics and the Indian Institute of Technology Bombay.

\subsection{Observing site location}

GIT is situated at the Indian Astronomical Observatory (IAO), located atop Mt. Saraswati, Digpa Ratsa Ri in Hanle, Ladakh. 
The observatory is located at latitude $32\degr46^{\prime}46^{\prime\prime}$ N and longitude $78\degr57^{\prime}51^{\prime\prime}$ E, at an altitude of $\sim$4500~m above the sea level. The IAO site has $\sim$ 190 clear photometric nights per year \citep{2002ASPC..266..424C}. IAO has meagre precipitation of $<7$~mm annually, thanks to Himalayan mountains that cast a rain shadow on the observatory. With low temperature, low humidity, a typical seeing of $<1^{\arcsec}$~\citep{2002ASPC..266..424C} and median sky brightness of $\mu_\mathrm{V} = 21.28~ \mathrm{mag/arcsec^{2}}$~\citep{2008BASI...36..111S}, IAO is one of the best sites for optical astronomy in India. The observatory has its weather station accessible by all telescopes present at the observatory. However, in the interest of an additional level of safety, the weather station is not directly integrated into the telescope system, and the dome closure/opening is handled by the staff at other telescopes of the observatory. The observatory has an engineering team that regularly handles the maintenance of the telescope system. The observatory is remotely accessible through a satellite link from the CREST campus of the Indian Institute of Astrophysics, Bangalore.

\begin{figure}
\centering
\includegraphics[width=\columnwidth]{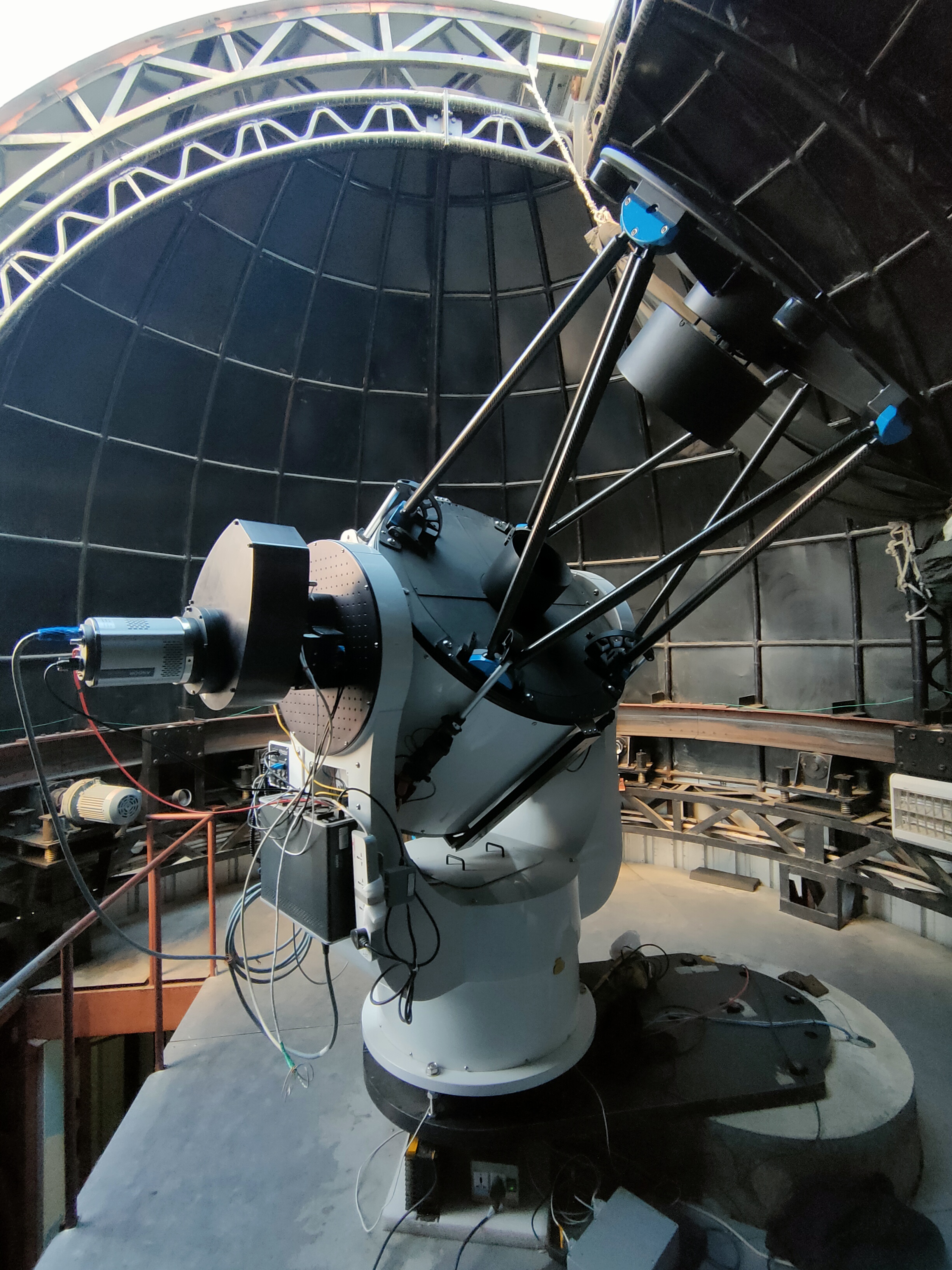}
\caption{GROWTH-India Telescope inside the enclosure. The Andor camera is shown mounted on the telescope.}  
\label{fig:git}
\end{figure}

\subsection{Telescope}
The telescope is a Planewave 0.7~m diameter primary mirror telescope with a Corrected Dall-Kirkham (CDK) optical design\footnote{\url{https://planewave.com/product/cdk700-0-7m-cdk-telescope-system/}} (Figure~\ref{fig:git}). With a focal ratio of f$/6.5$ and focal length of 4540~mm, the telescope provides a 70~mm image circle, corresponding to a $\sim0.86$~degree diameter circle on the sky. The Nasmyth focus design and dual truss structure make it possible to mount two instruments simultaneously and switch between them with minimal overheads.
This CDK telescope is driven by integrated direct-drive motors, providing zero backlashes, zero periodic error, and requires minimal maintenance. The telescope can reach slewing speed of up to 50 degrees/s enabling it to track fast-moving satellites and near-earth objects. Further, the telescope has a very good pointing accuracy of $10\arcsec$ and tracking accuracy of $<1\arcsec$ for a typical 300~s exposure. For all practical purposes we limit the longest exposure to 10 minutes. The full specifications of the telescope are listed in Table~\ref{tab:telcamspec}.
\begin{table*}
    \centering
        \begin{tabular}{|l|p{6cm}|}
       \hline
       \multicolumn{2}{|c|}{\textbf{Telescope Specifications}}\\
        \hline
        Optical Design &  Corrected Dall-Kirkham (CDK) \\
        \hline
        Mount type &  Alt-Az mount\\
        \hline
        Aperture & 	700~mm (27.56~inch) \\
        \hline
        Focal Length & 4540~mm \\
        \hline
        Focal ratio	 & 6.5 \\
        \hline
        Central Obstruction  & 47\% of Mirror Diameter \\
        \hline
        Optical Tube  &  Dual truss structure with Nasmyth focus \\
        \hline
        Back focus & 309~mm from Mounting Surface \\
        \hline
        Optical Performance & 1.8~micron RMS spots \\
        \hline
        Field of View & 70~mm image circle (0.86 degrees) \\
        \hline
        Pointing Accuracy &	 10\arcsec RMS \\
        \hline
        Maximum Altitude for pointing & 88 degrees $\sim$ 6.5 deg$^2$zenith blind spot \\
        \hline
        Pointing Precision &  2\arcsec \\
        \hline  
        Tracking Accuracy &  $< 1$\arcsec error over 10 minute period \\
        \hline
        Field De-Rotator Accuracy &	3 microns of peak to peak error at 35~mm off-axis over 1 hour of tracking \\
        \hline
        Primary Optical Diameter & 	700~mm (27.56~inch) \\
        \hline
        Secondary Optical Diameter &	312.4~mm \\
        \hline
        Tertiary Optical Major Diameter	 & 152.4~mm \\
        \hline
        Mirror Material (All)  & Fused silica (quartz) \\

        \hline
        \hline
        \multicolumn{2}{|c|}{\textbf{Camera Specifications}}\\
        \hline
        Sensor Type & CCD230XL-84 midband AR coating \\
        \hline
        Pixels & 4096 (H) $\times$ 4108 (V) \\
        \hline
        Pixel size & 15 $\times$ 15~$\mu$m \\
        \hline
        Pixel scale & 0.676\arcsec/pix \\ 
        \hline
        Image area & 61.4~mm $\times$ 61.4~mm \\ 
        \hline
        System Window transmission & $> 98~\%$ \\
        \hline
        Cooling & $ -55^{\circ}$C air TE cooled, up to $-100^{\circ}$C deep TE cooled. Operated at $\textbf{-40}^{\circ}$C \\
        \hline
        Well depth & 1,50,000$~\elec$ \\
        \hline
        Readout rates & 0.1, 1, \textbf{2}, 4~MHz \\
        \hline
        Readout noise &  3.8, 8.5, \textbf{12.0}, 23.0$~\elec$ \\
        \hline
        Dark current @ $-55^\circ$ & 0.001~$\elec$/pixel/s\\ 
        \hline
        Peak quantum efficiency &  $> 95\%$ \\
        \hline
        Linearity &  $ > 99\%$ \\
        \hline
        Timestamp &  IRIG-B GPS with 10~ms resolution \\
        \hline
        Gain  & $1 \times 1.04~\elec/mathrm{/ADU}$ \\
        \hline
        Depth (300s exposure r$^\prime$-band) & $\sim$~20.5 mag \\
        \hline
         \end{tabular}
    \caption{Planewave CDK-700 telescope (upper half table) and Andor iKon XL-230 Camera specifications (lower half table). Default values are indicated by bold text.}
    \label{tab:telcamspec}
\end{table*}

\subsection{Instruments}
GIT uses an Andor XL-230-84 CCD as its primary instrument. In addition, the system's performance has been tested with APOGEE Alta U32 and SBIG STF 8300EN instruments which were used on the system during commissioning phase and during the unavailability of the primary camera due to technical failures. The following section provides details of all cameras utilised on GIT to date. The cameras are operated through automated command-line scripts to perform all operations.

\subsubsection{Primary instrument: Andor} \label{sec:andor}
The primary instrument for GIT is an Andor XL-230 camera, which is also in use at present. This 4096 $\times$ 4108 pixels camera, in combination with CDK 700 telescope, provides an effective field of view of $49\arcmin$ radius circle. This back-illuminated camera has high quantum efficiency (QE) of $> 95\%$. The built-in thermo-electric (TE) system and air cooling can lower the operating temperatures to 60\degr~ below ambient. The operating temperatures can be lowered further with water cooling, but this is not used at GIT due to low ambient winter temperatures and robustness requirements for an autonomous telescope. We chose the CCD with 15 $\mu m$ pixel size to sample the 
point spread function (PSF) properly, providing a pixel scale of $0.676^{\arcsec}$. The CCD has a full-well depth of 1,50,000~$\elec$, providing a large dynamic range. With linearity of $>99\%$ and very low dark current ($\sim 0.0024~\mathrm{ ADUs/pix/s}$), the CCD is well suited to long exposures without significant non-uniformity in images. 
The CCD supports a wide range of readout speeds, from a 10~kHz readout giving a noise of just 3.5~$\elec$ to a fast 4~MHz mode that allows a full-frame readout in just 4~s, at the cost of a higher 23.4~$\elec$ readout noise. In practice, we typically use a single-port readout at 2~MHz. User-definable binning makes the CCD versatile to use in desired science cases. Detailed specifications of this Andor camera are listed in Table~\ref{tab:telcamspec}, while more details from the vendor are available at \url{https://andor.oxinst.com/products/ikon-xl-and-ikon-large-ccd-series/ikon-xl-230}.

\subsubsection{Apogee}
An Apogee Alta U32 camera with a 3-megapixel Kodak Blue Plus sensor was used at GIT from August 2020 -- March 2021. The camera has a rectangular image layout with 2184 $\times$ 1472 pixels, where each square pixel has a 6.8-micron size (0.346\arcsec on the sky). The camera has an imaging area of $14.8~\times~10.0$~mm that provides an FoV of $11.17\arcmin~\times~7.52\arcmin$. The camera's full well capacity is 55000~$\elec$
a peak QE (610~nm) $\sim 85\%$. The camera features TE cooling with forced air, which enables it to cool to a maximum of $50^{\circ}$C below ambient temperature with a temperature stability of $\pm 0.1^{\circ}$C. This camera can expose in the range of 30~ms to 183~min with 2.56-microsecond increments. The system is read out at 1~MHz with a low readout noise of 8$\elec$. 

\subsubsection{SBIG STF-8300EN}
Another camera used at GIT is an SBIG STF-8300EN camera with a Kodak KAF-8300 CCD sensor with a 3326 $\times$ 2504 pixel array. Each pixel is 5.4~$\mu$m square in size, providing a plate-scale of $0.24\arcsec$. Its 8.3 million pixels camera gives an imaging area of 17.96 mm $\times$ 13.52 mm. The camera has a high readout speed, enabling it to read the entire frame in $\sim 1$~s with a typical readout noise of 9.3~$\elec$. The even illuminated shutter helps in taking exposures as short as 0.1~s. On the higher end, an hour-long exposure can be obtained. This SBIG camera has a higher dark current of 0.15~$\elec$/pix/s despite an ability to cool up to 40--45\degr C below ambient temperature. Compared to our other two cameras, this camera has a lower full well capacity (25,000~$\elec$) and QE (56$\%$).

\begin{figure}
\centering
\includegraphics[width=8cm]{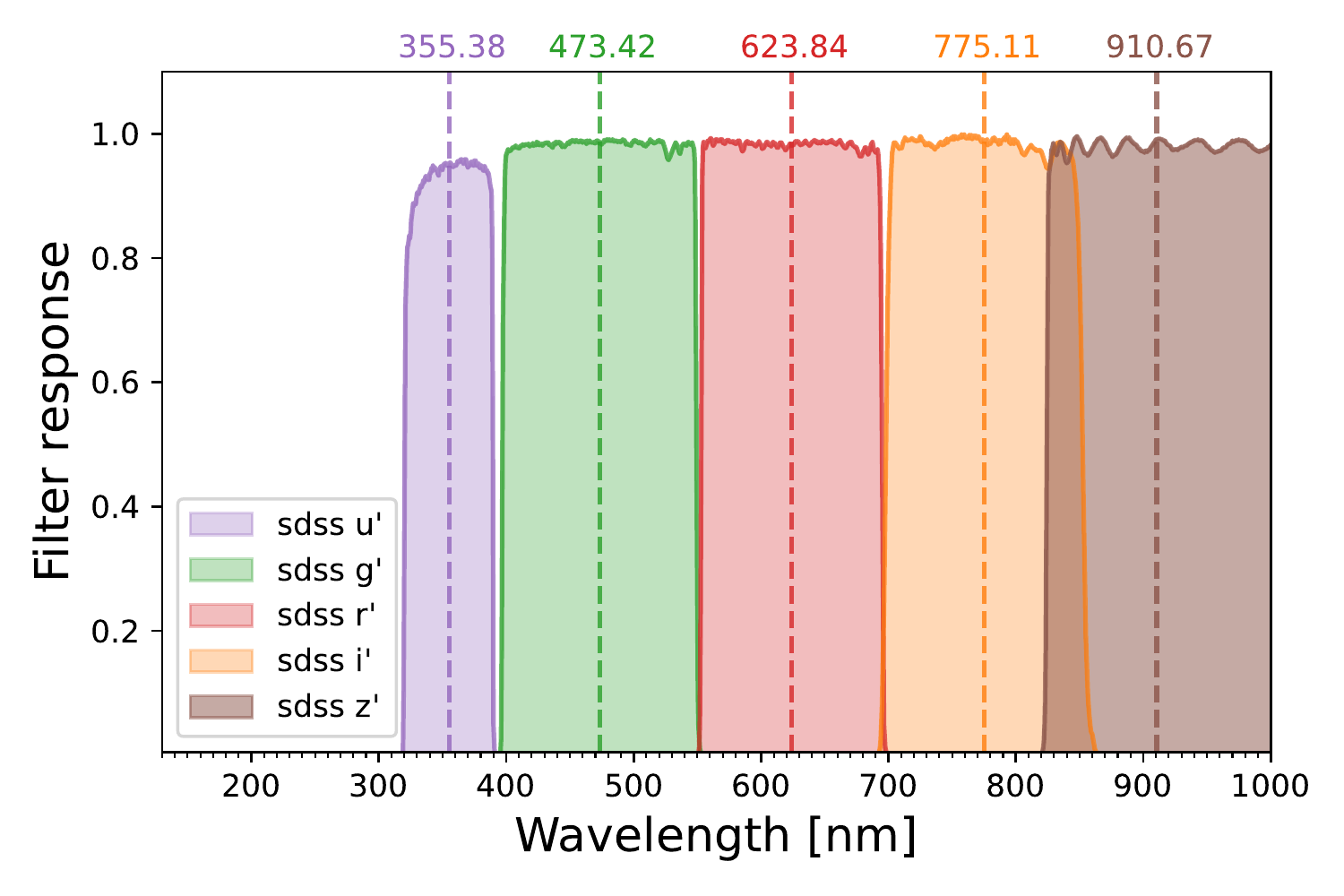}
\caption{Filter response curves for SDSS prime filter set used at GIT. The effective wavelengths for each filter are shown by a vertical dashed lines and their values are quoted at the top of the plot.}
\label{fig:filterresponse}
\end{figure}

\subsection{Filter system}
SDSS u$^\prime$ g$^\prime$ r$^\prime$ i$^\prime$ z$^\prime$ filters (Figure~\ref{fig:filterresponse}) are integrated into a filter wheel assembly design. The closed-loop control system of the filter wheel provides precise position and velocity control. The system is realised with a PSoC5LP micro-controller unit (MCU) based on system-on-chip (SoC) architecture. The filters are arranged in a wheel that can rotate in both clockwise and anti-clockwise directions to minimise the filter change time while changing extreme position filters. A python-based script communicates with the filter wheel via an RS232 interface to perform all operations like the status of the wheel, homing position, current filter position, filter change operations, etc. The typical filter change time among adjacent filters is 2.30~s.

\section{Observing with GIT}\label{sec:obs}

GIT operates in a default automated mode for nightly observations. The entire telescope assembly operations are controlled by two PCs that run on the Linux and Windows operating systems. The core component is the Linux PC, which runs the custom python-based operations software for GIT. This includes scheduling, controlling the camera, filter wheel, data storage and archiving, listening for Target of Opportunity commands, error reporting, etc. The telescope and dome are controlled by the Windows PC that hosts the softwares for the telescope and dome controls. These softwares open a socket to send and receive commands for various operations of the telescope and dome, thus controlling them via communicating over Ethernet without any manual intervention. For higher-level operations like rebuilding the pointing model, the observer directly accesses the softwares through the windows PC at the telescope site. We discuss key components of the control software in the following sections.

\subsection{Observation Scheduling}
\begin{figure*}
\centering
\includegraphics[width=0.9\textwidth]{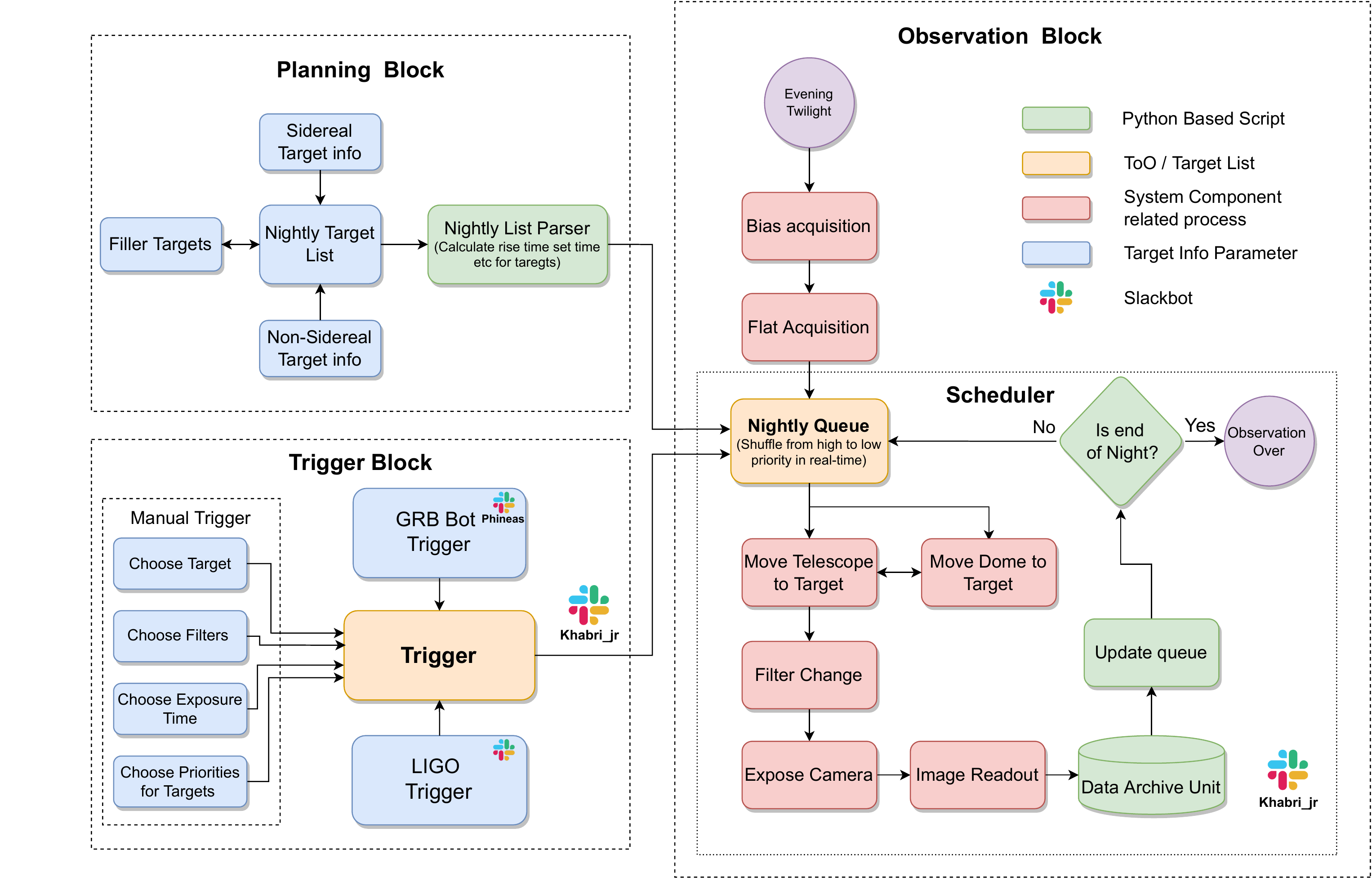}
\caption{Observation procedure of GIT. The procedure is broadly divided into three blocks: `Planning' of targets (Top left block), `Observations' which includes scheduling (Right block) and adding additional targets using `Triggers' (Lower left block). Observations are automated and are monitored by slack bots.}
\label{fig:flowchart}
\end{figure*}

The default robotic observation mode of GIT is the queue-based observation mode. Observation scheduling of GIT is illustrated in Figure~\ref{fig:flowchart}. An object list is prepared before starting the nightly observations, containing all relevant information for selected targets. This list is parsed to the scheduling computer, which orders the targets based on their priority assigned by the observer. The observer chooses this priority by taking a particular target's set time and scientific importance into account. The control computer automatically takes the information from the object list for any particular object and performs observations. New targets like Gamma-Ray Bursts (GRBs) can be manually or automatically added to the queue at any time by using specialised remote scripts. This is particularly important for rapid response to GRBs etc. Once a target is added to the queue, it is processed as per the assigned priority like any other target in the queue. This means that the ongoing image sequence is finished before switching to any such new target. 

\subsection{Calibration image acquisition}
Nightly observations begin with a series of calibration exposures that are used for processing all science images acquired during that night. 
\subsubsection{Bias frames}
Bias frames are zero-second readouts to measure the unwanted electronic signal. Bias frames for our primary Andor camera are stable throughout the night. Therefore, we acquire these frames at the start and end of the night. The CCD is cooled to $-40^\circ$C, which helps stabilise the bias frames despite ambient temperature fluctuations.

\subsubsection{Dark frames}
A dark frame measures the thermal noise of the camera, which is acquired by taking images with the shutter closed to avoid external photons contributing to the image counts. We minimise the dark current by cooling the camera to $-40^\circ$C. As mentioned in \S\ref{sec:andor}, the Andor CCD has a very low dark current. To verify the same, we acquired the dark frames at a fixed temperature at various exposure times ranging from 150~s - 700~s in steps of 50--100~s. Using a linear fit to the data, we obtained a typical dark current of 0.0052~ADUs/pix/s at $-40^\circ$C CCD temperature, broadly consistent with the camera datasheet. This amount of dark current does not impact the science exposure. Hence, the dark frame acquisition can be omitted for exposure $<600$~s.

\subsubsection{Flat frames}
Twilight flat frames are used in GIT to correct the pixel response of CCD. 
A python-based code automatically acquires well-exposed flat frames for all filters during the evening twilight. $\rm{u}^{\prime}$ and $\rm{z}^{\prime}$ band flats are acquired first as they require brighter sky conditions due to high extinction in these bands. These are followed by $\rm{g}^{\prime}$, $\rm{r}^{\prime}$ and $\rm{i}^{\prime}$ band exposures. The code checks the count levels in the images and eliminates any saturated or under-exposed flat images. It adjusts the exposure time for subsequent flat frames to maintain median counts in the range of 30,000--45,0000.

\subsection{Non-sidereal mode observations}
Non-sidereal observing has been integrated with the main scheduler of GIT to streamline observations of moving objects. With GIT, we perform two types of objects in non-sidereal mode: 1) confirmation of newly discovered Near-Earth Objects (NEOs) by GROWTH partners and other observers,
2) follow-up observations of specific targets with confirmed orbits. 

Due to the significant sky plane motion of these objects with respect to fixed background stars, the coordinates of these targets keep on changing with time. For objects confirmed by Minor Planet Centre (MPC), a list of targets is prepared as per scientific interests. This list resides at IIT Bombay and comprises designations and tentative observation times of the objects. Daily at 5~PM local time, the coordinates of these non-sidereal targets are updated automatically by querying the MPC database. After querying the coordinates, a non-sidereal nightly file is generated as per the required format of the GIT scheduler, which is then synced automatically to the system at the observatory in Hanle, where it gets combined with the sidereal target list. Since the non-sidereal targets may have a significant sky-plane motion, we approximate the coordinates of the targets at the time of observation by linearly extrapolating the coordinates in the nightly file. For potential Near-Earth Object (NEO) discovered on the same night, the non-sidereal observations can be triggered on GIT in real-time, thus enabling active follow-up of these candidates.

\subsection{Helper bots}\label{subsec:bots}
The GIT system is integrated with several helper ``bots'' that monitor various operational parameters. These bots perform various tasks like automatically triggering telescopes for ToO observations, monitoring observation as well as data processing, and summarising the last night's observations in the form of a nightly report. These reports are then posted on slack\footnote{\url{https://slack.com/intl/en-in/}.} groups for human inspection. In critical situations, telephonic / text message alerts can also be sent to designated users.

``\sw{Khabri\_Jr}": GIT observations are monitored by the \sw{Khabri\_Jr} slack bot. This bot updates the observers about a night's observation plan and sends near real-time telescope observation updates via slack. In case of instrument failures, the scheduling system can debug first-order errors. However, in certain cases, when the debugger fails to solve the problem, the slack bot can identify the errors in the real-time observation log and notify designated users by a telephone call through a ``Twilio'' API\footnote{\url{https://www.twilio.com/voice}} so that debugging can be handled manually.

``\sw{Phineas}": This slack bot continuously looks for the GRB alerts on  Gamma-ray Coordinates Network/Transient Astronomy Network (GCN/TAN) server\footnote{\url{https://gcn.gsfc.nasa.gov/about.html}}. Upon receiving a trigger alert, the bot checks it against triggering criteria for GIT and generates a trigger if these criteria are met. The bot sends the generated trigger to the telescope scheduling computer to add it to the main queue. A similar bot looks for Gravitational Wave (GW) event candidates during the observing run of the advanced GW detectors (LIGO, Virgo, Kagra) and notifies designated users through a phone call. For future runs, this script is also being upgraded to create an observing schedule and start observations in parallel to the user notifications.

``\sw{Khabri}": The GIT automated data reduction pipeline is integrated with Khabri (messenger) bot, which tracks data download and reduction processes. Near real-time processing, updates are sent to slack during data processing. On completion of data reduction, the bot generates a summary report which contains information about the observed targets, time of observations, filter information, reduction status, and GIT observation performance statistics for last night's observations. At the end of each night's observation, the bot generates nightly statistics plots for images' seeing and depth variation during the night as a part of the summary report.

``\sw{Gitty}": This bot is designed to send data to relevant observers automatically. After the generation of the nightly summary report, the bot filters out data under different proposals and syncs the processed data to the necessary destinations for observers to access in minimal time.

\section{Data processing}\label{sec:dataproc}

The automated scripts regularly download data from the observatory without any human intervention. Once downloaded, this data is processed by a fully automated data reduction pipeline and archived.

\subsection{Data transfer}
All the raw data are stored at the data storage unit at IAO and need to be transferred to IIT Bombay for processing. These data transfers are executed through the CREST campus of IIA, which is connected to IAO through a satellite link. Data are downloaded to the data relay unit at CREST in the daytime for regular observations. However, we download data in real-time via automated download scripts for important time-constrained observation. A network-attached storage (NAS) at the CREST also serves as a redundant backup of all raw data. The typical GIT raw data load for a single night is $\sim 3$ Gigabytes. Data files arrive at CREST and are immediately copied to IIT Bombay over the internet for further processing. A NAS system at IIT Bombay archives the raw as well as the processed data that can be accessed by an internal query interface. All the data remain proprietary.

\subsection{Image reduction}
The image reduction and processing pipeline handles all tasks required for converting the observations into values ready for interpretation.

\subsubsection{Pre-Processing}\label{sec:preprocessing}
To obtain the flux from an astronomical source, all raw GIT exposures taken with a CCD imager have to go through a set of generic steps before further image-specific processing. These include bias-subtraction, flat-fielding using the bias and flat frames obtained on the same night, and cosmic-ray correction using \sw{Astro-SCRAPPY}~\citep{2019ascl.soft07032M} package. The final pre-processing step is to calculate the astrometric solution to map pixel coordinates to sky coordinates, using the \sw{solve-field} offline package of astrometry.net \citep{Lang_2010}.

\subsubsection{Photometry}
Upon pre-processing, the images are processed via standard image processing routines. All point sources are extracted using the \sw{SExtractor}~\citep{Bertin_1996}. These sources are cross-matched with PanSTARRS DR1~\citep{chambers2016} catalogue through a \sw{VizieR} query to obtain the zero-points of the image. The brightness of sources is estimated using PSF photometry on the source of interest. To perform PSF photometry, we obtain the PSF profile of the sources in the image using the \sw{PSFEx} package~\citep{bertin11} and convolve it with the source image to estimate the source flux. The instrumental magnitude thus obtained is corrected using an earlier estimated zero-point to obtain the standard magnitude of the source.

\subsubsection{Image subtraction}
In some instances, the source of interest lies embedded inside their host galaxies, which makes it difficult to perform accurate photometry on these objects due to host contamination. Broadly, searching for new transients in the image also requires image subtraction. Such specific images of interest (identified by a header flag) go through another set of data reduction steps in the form of image subtraction and candidate search.

Image subtraction is a numeric method to subtract a deeper reference image from the science image of interest. The GIT Image subtraction pipeline is based on the ZOGY algorithm \citep{2016ApJ...830...27Z}. Owing to the relatively large FoV, we first divide the processed image into smaller cutouts to reduce the impact of PSF and background variation across the image. Corresponding Pan-STARRS DR1 (PS1) image cutouts are downloaded as reference images using \sw{panstamps}\footnote{\url{https://panstamps.readthedocs.io/}}. Astrometric uncertainties in science and reference images are reduced by a cross match of positions with Gaia Data Release 2 ~\citep[Gaia DR2]{Brown2018} catalogue. \sw{SWarp}~\citep{SWARP_Bertin} is used to re-sample both science and reference images on the same astrometric grid to align their position angles. The reference image is scaled in flux to the science image flux level based on the local ratio of pixel scales and the zero-point of images. Finally, the re-sampled flux-matched science and reference images, the PSF model, the RMS images, and the astrometric uncertainties are used as input to \sw{pyzogy} \citep{pyzogy_2017} to obtain the difference image and to correspond to the statistics image (score image).

\subsubsection{Transient search}
A vast majority of GIT observations are performed in ``targeted mode'', typically having a single source of interest for photometry. However, for certain science goals like searching for poorly localised GRB afterglows and GW events counterpart searches, GIT undertakes ``tiled mode'' observations acquiring multiple contiguous images covering the sky region to which the transient has been coarsely localised. We run the image subtraction pipeline on these tiled mode observations. A subsequent pipeline with human-in-the-loop vetting is used for identifying transients in the subtracted images. We provide a very short overview here, and for details, refer the reader to Kumar et al. (in prep). Transient candidates are identified by locating the position of local peaks in the ``score'' image with $s_\mathrm{corr}$ value $>$ 5. Among these, most of these detected candidates ($> 99\%$) are spurious and filtered out by the candidates vetting pipeline. Finally, the remaining candidates are scanned by humans for verification. The interesting candidates thus obtained (if any) are further followed up to measure their evolution in time.

\subsubsection{Processing non-sidereal data}
The primary objective of non-sidereal observations with GIT is to report accurate astrometry and photometry of NEOs to MPC to reduce their orbital uncertainties. Images acquired in non-sidereal mode show all stars as streaks, with the target object ideally seen as a point source. We have developed another special pipeline to process such data: ``Astrometry with Streaking Stars'' \citep[\sw{Astreaks};][]{2021EPSC...15..378S}. \sw{Astreaks} identifies all the streaks in the image, generates an equivalent synthetic image with point sources and solves it for astrometry using standard astrometry packages\footnote{Discussion of Astreaks implementation \url{https://sites.google.com/view/growthindia/results/astreaks}}. Astrometry obtained using this pipeline results in sub-arcsecond $Observed - Computed$ residuals. Along with observations of Fast Moving objects, we have also done follow-ups of multiple cometary outburst events. The aperture photometry of comets with centring is fully automated, given the aperture size. The GIT pipeline is also equipped with track-and-stack software designed to stack the data of moving targets, thus helping with better analysis of cometary features by improving the SNR. The detailed working of \sw{Astreaks} is discussed in K. Sharma et al. (in prep).

\subsection{Observation reports and data archiving}
GIT pipeline has been designed to produce standardised data products. The images obtained after the entire processing have standard headers and do not require any additional knowledge about GIT by users. Data processing is logged and reported on slack in real-time, which helps in debugging the unprecedented processing errors and tracking live-time processing progress. All raw and processed data are archived in a Network Attached Storage (NAS) at IIT Bombay. Key data properties are also stored in a local queryable database which is populated right after the completion of processing of each image.

\begin{figure}
\centering
\includegraphics[width=8cm]{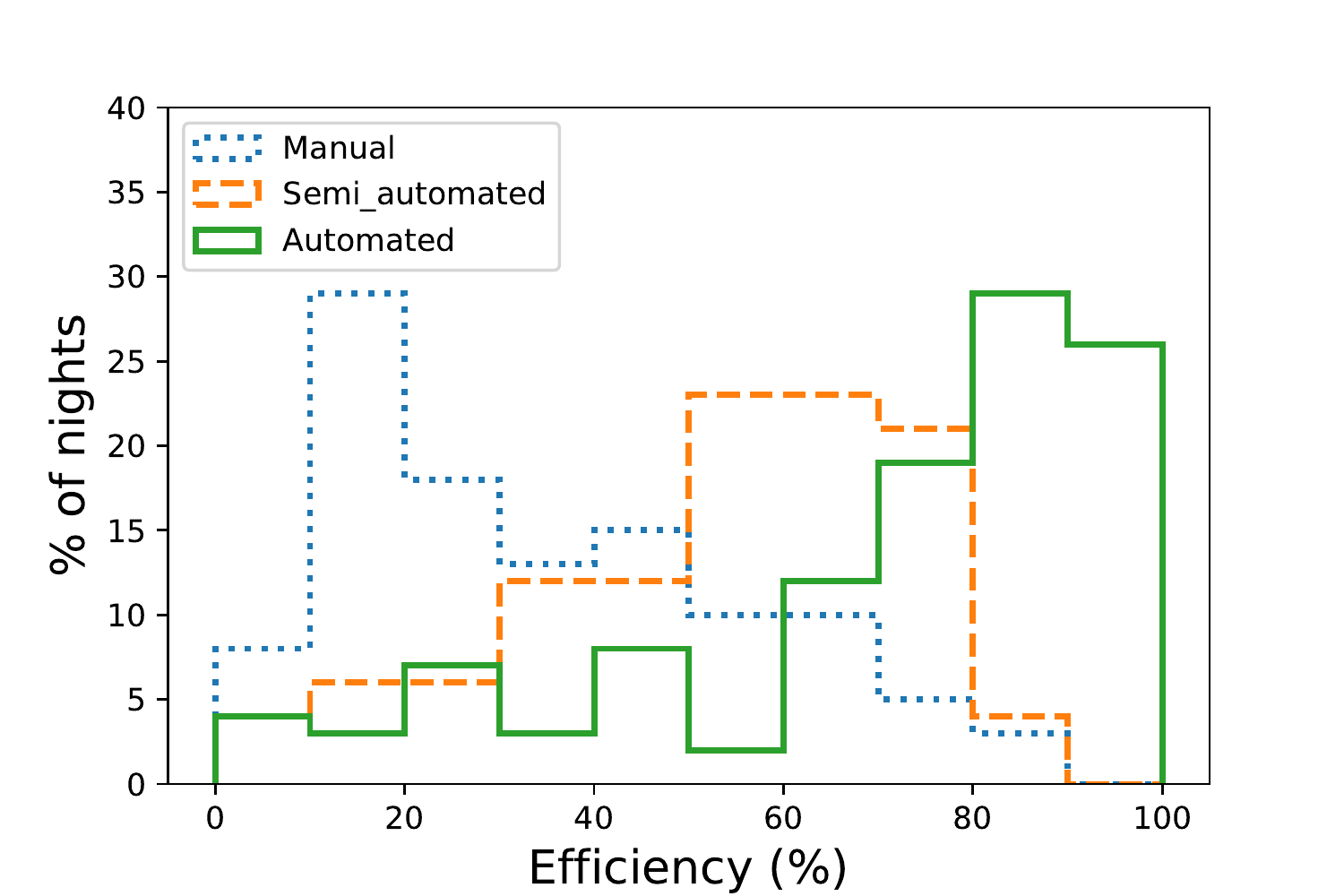}
\caption{GROWTH-India Telescope efficiencies during each phase of operations. Three histograms show the efficiencies for manual mode (blue), semi-automated mode (orange) and automated mode (green). The system is currently running in automated mode with a typical efficiency of $\sim 80\%$, which is a noticeable improvement over the manual and semi-automated mode of observations. The low effacing tail in automated mode is due to unsuitable weather conditions and instrument failures.}
\label{fig:eff}
\end{figure}

\section{System performance}\label{sec:perf}

\subsection{On-sky efficiency}
GIT received its first light during the commissioning phase in June 2018. At commissioning, the telescope and camera system were operated in manual mode, with remote and local operators sharing responsibilities. Since then, the system has been upgraded in a phased manner. In Feb 2019, the system was upgraded into the semi-automated mode, where scripts managed the bulk of operations. However, a few tasks like dome operations, calibration frames acquisition, and system monitoring were handled by operators at the observatory. In Sept. 2021, GIT was upgraded to a fully automated mode.

In the course of typical operations, the control software handles all responsibilities, including acquiring calibration images, calculating the observing schedule from the master target list, filter changes, target acquisition and tracking, operating the dome, filter wheel, camera, etc. For absolute safety, the control software does not have the capability to operate the dome or telescope shutters: these are the only operations that require the physical presence of the observing assistant at the start and end of the observing program. 
The control software monitors various diagnostic parameters and is capable of first-order debugging. If the system encounters any problems which it cannot fix, it alerts observers via messages or phone calls through the various bots (\S\ref{subsec:bots}).

We quantify the telescope performance in two efficiency parameters. The first is the ``on-sky efficiency'': the fraction of the night (defined by nautical twilight) that was spent acquiring target images. Thus, even necessary operations like readout and slew are considered as extraneous in this definition. GIT currently operates with typical nightly efficiencies of $\gtrsim 80\%$. 
We also define an ``effective efficiency'', where we include the readout time as part of the exposure. In this case, the ``effective inefficiency'' arises from weather, technical issues, or slews (which can be optimised to some extent). On good observing nights, effective efficiencies as high as 94\% have been achieved (Figure~\ref{fig:eff}). In contrast, the median effective efficiency in manual mode operations was around 30\%. Part of this can be attributed to the upgrades and testing that were a key feature of manual operations. In semi-automated mode, median effective efficiency rose to $\sim 60\%$. The highest efficiencies in manual or semi-automated nights were attained during campaigns where we monitor a single target for most of the night. This drastically reduced the need for telescope or dome slews or any human intervention in general. Overall, the best efficiencies are obtained in fully automated observations. Figure~\ref{fig:eff} shows the clear increase in telescope utilisation efficiency from manual to semi-automated and finally to fully automated observations. The low-efficiency tails of the semi- and fully-automated observations arise from partial nights lost to weather problems and to occasional technical issues.

\subsection{Sensitivity}
GIT has good sensitivity, with a typical depth of $\sim$ 20.5--21 magnitudes in 10-minute g$^\prime$ and r$^\prime$ exposures. The typical full-width at half-maximum for point sources in GIT data is $\sim 3\arcsec$. Figure~\ref{fig:limmag} shows the 5-sigma limiting magnitudes of GIT images for various filters for data obtained between July 2021 to Dec 2021. The median limiting magnitude is 20.44 in g$^\prime$ and 20.49 r$^\prime$ bands for data obtained on $\pm 5$ days of the new moon. u$^\prime$ and z$^\prime$ band images are typically shallower due to a combination of attenuation, reflectivity, camera sensitivity, and background. The faintest magnitude achieved with a 120-minute co-add in the r$^\prime$ band was $\sim 22.9$.
In all bands, data obtained within $\pm 5$ days of the full moon are relatively shallower by up to two magnitudes due to the brighter sky background. Very shallow images in Figure~\ref{fig:limmag}, with a limiting magnitude around 16, are ones acquired under very poor observing conditions, like partial clouds.

\subsection{Photometric accuracy}\label{sec:accuracy}
\begin{figure}
        \fig{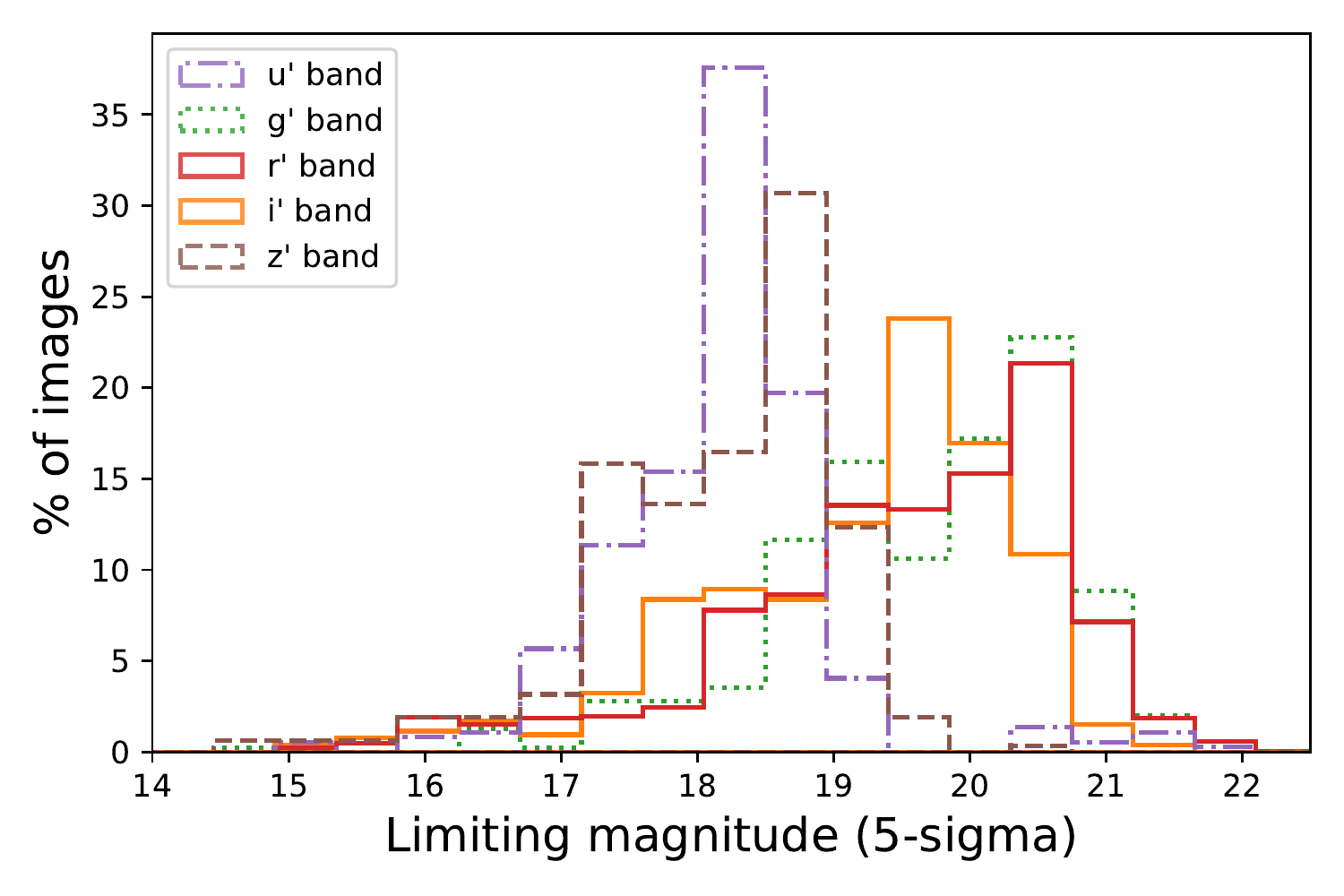}{\columnwidth}{(a)} \label{fig:limmag}\\
        \fig{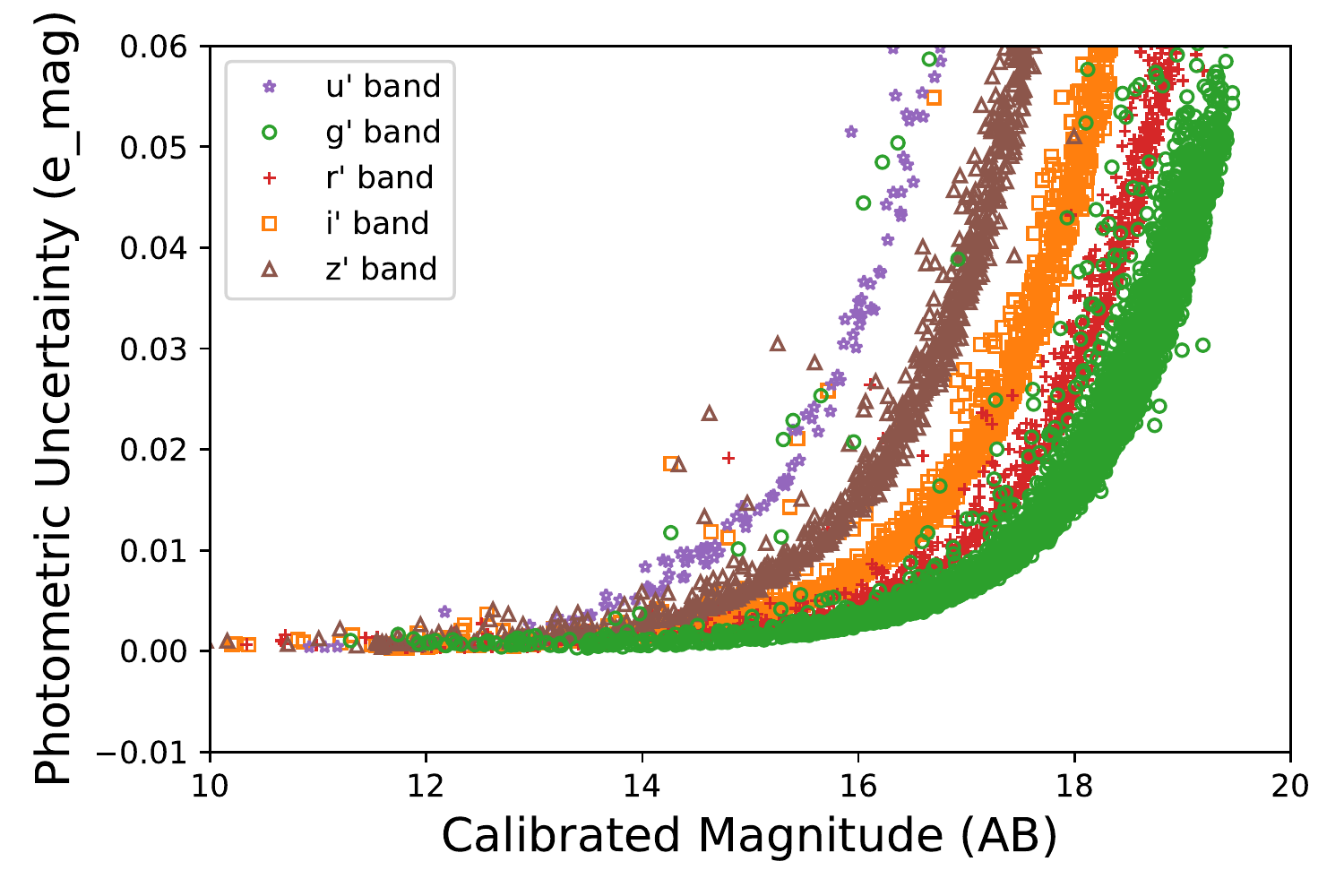}{\columnwidth}{(b)} \label{fig:photuncertn}\\
        \fig{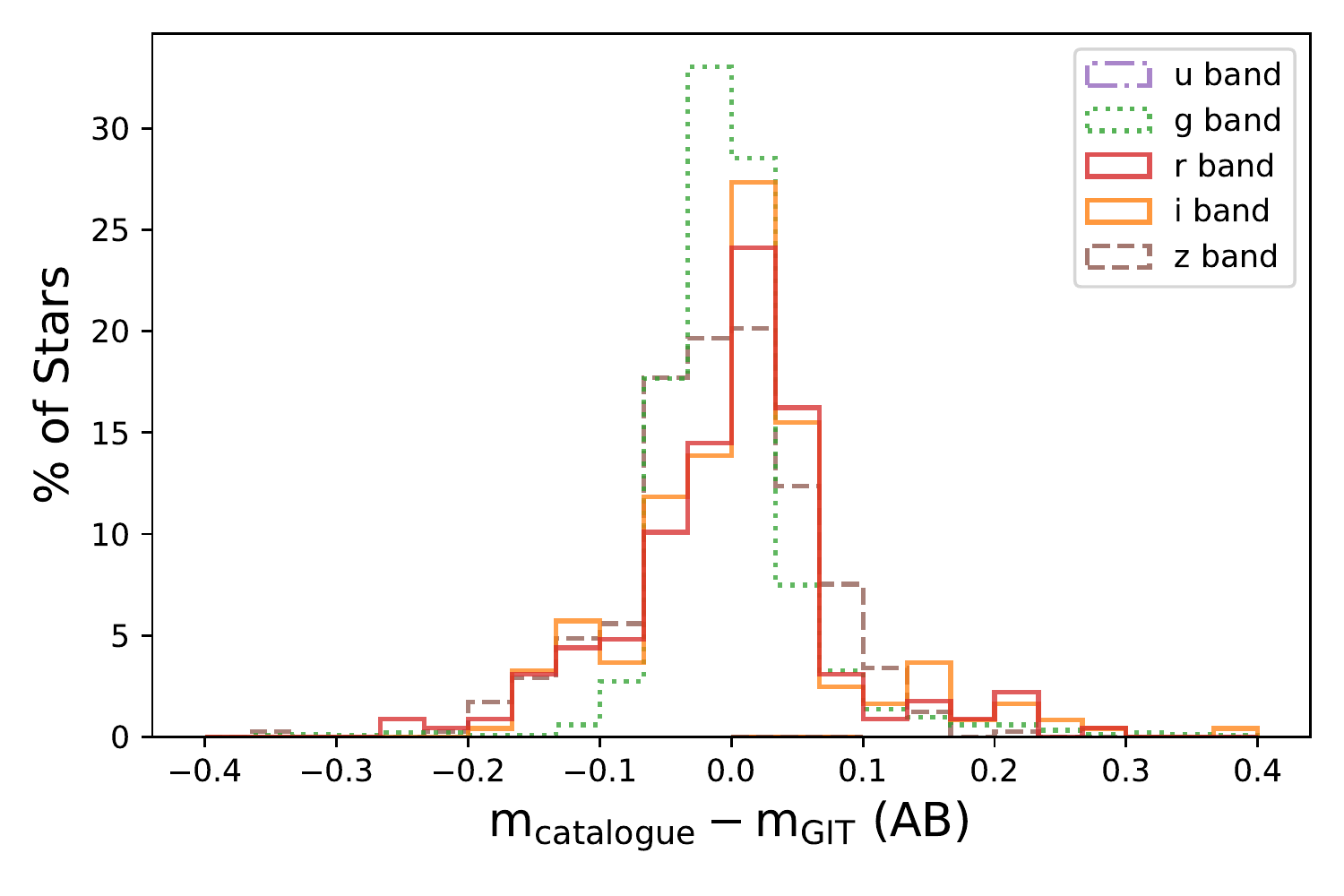}{\columnwidth}{(c)} \label{fig:photacc}\\
        \caption{(a) Limiting magnitude variation of GIT-Andor camera in SDSS $\mathrm{u}^\prime \mathrm{g}^\prime \mathrm{r}^\prime \mathrm{i}^\prime \mathrm{z}^\prime$ bands for 300-sec exposures. The $\mathrm{g}^\prime$ and $\mathrm{r}^\prime$ band images are deeper as compared to other bands due to low extinction and has typical depth of 20.5 mag for 300~s exposure. (b) Photometric uncertainty for sources in various bands, as a function of source magnitude. (c) Photometric accuracy of GIT pipeline, measured as the difference between GIT and catalogue magnitudes for all sources discussed in \S\ref{sec:accuracy}.
        \label{fig:photaccuracy}}
        \end{figure}

GIT images are calibrated by cross-matching the stars from the same field against the Pan-STARRS Data Release 1 ~\citep[DR1;][]{chambers2016} for the g$^\prime$ r$^\prime$ i$^\prime$ z$^\prime$ bands, and the SDSS DR12 catalogue~\citep{2015ApJS..219...12A} for u$^\prime$ data. To check the reliability of photometric calibration for g$^\prime$ r$^\prime$ i$^\prime$ z$^\prime$ bands, we selected observations obtained on several dark nights (new moon $\pm5$~days), with airmass $<1.1$ and exposure time 300~s. For the u$^\prime$ band, we used a stacked image of 600~s. In all bands, we limited the analysis to stars with a signal-to-noise (S/N) ratio $> 10$. The photometric accuracy for various bands is plotted against the magnitudes of stars in Figure~\ref{fig:photaccuracy}.
Further, we derived photometric transformation equations between the GIT filter system and SDSS / Pan-STARRS filters. These transformation equations were obtained using a set of isolated stars in GIT images and reference catalogues. The transformation equations for Pan-STARRS and SDSS catalogues are listed below. 

\noindent Transformations from Pan-STARRS to GIT magnitudes:
\begin{eqnarray}
\mathrm{g^\prime_{GIT}} & = & \mathrm{g_{PS}} - 0.0051*(\mathrm{g_{PS} - r_{PS}}) + 0.0001,   \sigma = 0.0067 \label{eq:psgr} \nonumber \\
\mathrm{r^\prime_{GIT}} & = & \mathrm{r_{PS}} + 0.0016*(\mathrm{g_{PS} - r_{PS}}) - 0.0066,   \sigma = 0.0165 \label{eq:psrg} \nonumber \\
\mathrm{r^\prime_{GIT}} & = & \mathrm{r_{PS}} - 0.0085*(\mathrm{i_{PS} - r_{PS}}) - 0.0080,   \sigma = 0.0178 \label{eq:psri} \nonumber \\
\mathrm{i^\prime_{GIT}} & = & \mathrm{i_{PS}} + 0.0091*(\mathrm{i_{PS} - r_{PS}}) + 0.0023,   \sigma = 0.0133 \label{eq:psir} \nonumber \\
\mathrm{i^\prime_{GIT}} & = & \mathrm{i_{PS}} - 0.0158*(\mathrm{i_{PS} - z_{PS}}) - 0.0034,   \sigma = 0.0252 \label{eq:psiz}  \nonumber \\
\mathrm{z^\prime_{GIT}} & = & \mathrm{z_{PS}} - 0.0408*(\mathrm{i_{PS} - z_{PS}}) - 0.0100,  \sigma = 0.0159 \label{eq:psziz}\nonumber 
\end{eqnarray}

\noindent Transformations from SDSS to GIT magnitudes:
\begin{eqnarray*}
\mathrm{u^\prime_{GIT}} &=& \mathrm{u_{SDSS}} - 0.0168 * (\mathrm{g_{SDSS} - u_{SDSS}}) + 0.1044,  \sigma = 0.0565  \label{eq:sdssugu} \nonumber \\
\mathrm{g^\prime_{GIT}} &=& \mathrm{g_{SDSS}} - 0.1486 * (\mathrm{g_{SDSS} - r_{SDSS}}) + 0.0361,  \sigma = 0.0378 \label{eq:sdssgr} \nonumber \\
\mathrm{g^\prime_{GIT}} &=& \mathrm{g_{SDSS}} - 0.0719 * (\mathrm{u_{SDSS} - g_{SDSS}}) + 0.0383,  \sigma = 0.0389  \label{eq:sdssgu} \nonumber \\
\mathrm{r^\prime_{GIT}} &=& \mathrm{r_{SDSS}}- 0.0077 * (\mathrm{g_{SDSS} - r_{SDSS}}) + 0.0032,  \sigma = 0.0104 \label{eq:sdssrg}  \nonumber \\
\mathrm{r^\prime_{GIT}} &=& \mathrm{r_{SDSS}}+ 0.0021 * (\mathrm{i_{SDSS} - r_{SDSS}}) - 0.0023,  \sigma = 0.0071 \label{eq:sdssri}  \nonumber \\
\mathrm{i^\prime_{GIT}} &=& \mathrm{i_{SDSS}} + 0.0091 * (\mathrm{i_{SDSS} - r_{SDSS}}) + 0.0119,  \sigma = 0.0194 \label{eq:sdssir} \nonumber \\
\mathrm{i^\prime_{GIT}} &=& \mathrm{i_{SDSS}} - 0.0132 * (\mathrm{i_{SDSS} - z_{SDSS}}) - 0.0030,  \sigma = 0.0224 \label{eq:sdssiz}  \nonumber \\
\mathrm{z^\prime_{GIT}} &=& \mathrm{z_{SDSS}} - 0.0251 * (\mathrm{i_{SDSS} - z_{SDSS}}) + 0.0167,  \sigma = 0.0165 \label{eq:sdssziz}
\end{eqnarray*}

The higher colour term for the g$^\prime$ transformation is consistent with the PanSTARRs -- SDSS hyper-calibration \citep{Finkbeiner_2016}.

\section{Science results}\label{sec:science}

\begin{figure*}[t]
\includegraphics[width=\textwidth]{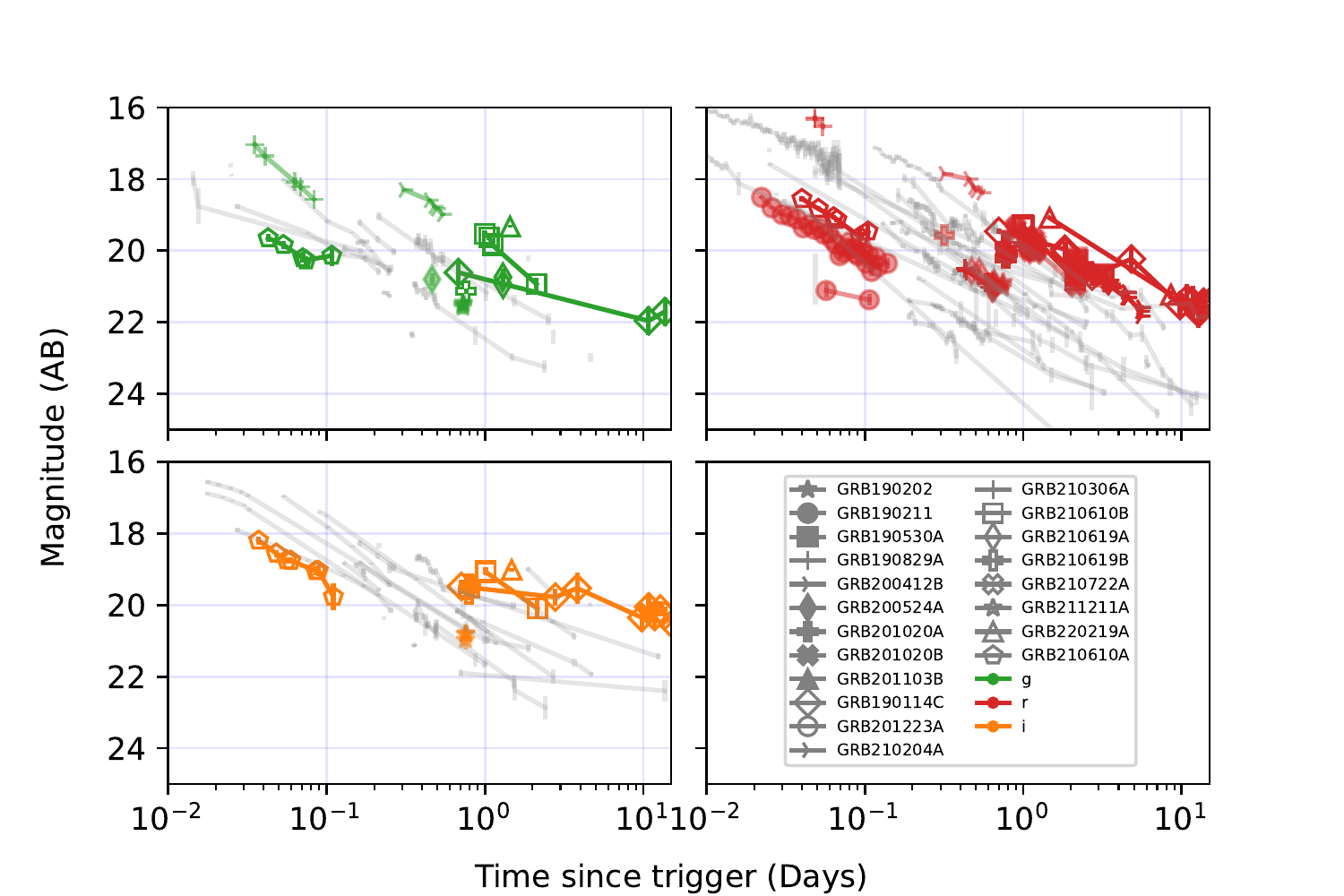}
\caption{Light curves of GRB optical afterglow detections by GIT. The light-curves of these afterglows in g$^\prime$ (top left panel), r$^\prime$ (top right panel) and i$^\prime$ (bottom left panel) bands has been plotted against Swift GRB afterglow population taken from \citet{2011ApJ...734...96K}. These afterglows span over the full range of Swift GRB afterglow population.}
\label{fig:GITgrbs}
\end{figure*}

The primary science goals for GIT are 1) The study of fast transients like gamma-ray bursts (GRBs) afterglows, electromagnetic counterparts to gravitational wave sources,  fast radio bursts, fast blue optical transients, etc. 2) Follow up of supernovae (SN) and Novae from very early to late phases, 3) The study and characterisation of solar system objects like Near Earth Asteroids, active asteroids, comets, etc.). In addition, the fourth goal of GIT is education ---  through courses, workshops, student training, etc. The last few years have been very productive for GIT in terms of the scientific outcomes, aided by its robotic capabilities. Since commissioning, GIT has followed up a good number of SN, Novae, GRB afterglows, EMGW candidates from GW merger events, near-Earth objects and comets. Further, we performed the transient search observation for suitable events like Fermi GRBs with small localisation uncertainty, and GW event tiled follow-up, etc. The results from these observations are summarised in the following section:

\subsection{Afterglows of Gamma-Ray Bursts}
GRBs have been the subject of great interest in the last two decades in astronomy. The afterglows of GRB arise due to the interaction of the jet with the surrounding medium at longer wavelengths than $\gamma$-rays. GRB optical afterglows are fast transients evolving on timescales of hours to days~\citep{2001grba.conf..300P}. GIT's autonomous operations and fast response time make these events a compelling science case for GIT. Since the first follow-up of such events in Dec 2018~\citep{2018GCN.23510....1S}, GIT has followed up $>70$ GRB events with 20 detections and put constraints on optical brightness of $>50$ events. Figure~\ref{fig:GITgrbs} shows the comparison of GRB afterglow detections followed up by GIT to the Swift GRB optical afterglow population~\citep{2011ApJ...734...96K}. We see that GIT has obtained dense sampling of several GRBs \citep[see for instance][]{GRB210204A-harsh-paper}, and that the afterglow candidates studied by GIT show similar fading characteristics as the broader sample of \emph{Swift} GRBs. 

\begin{figure*}[ht]
\centering
\includegraphics[width = \textwidth]{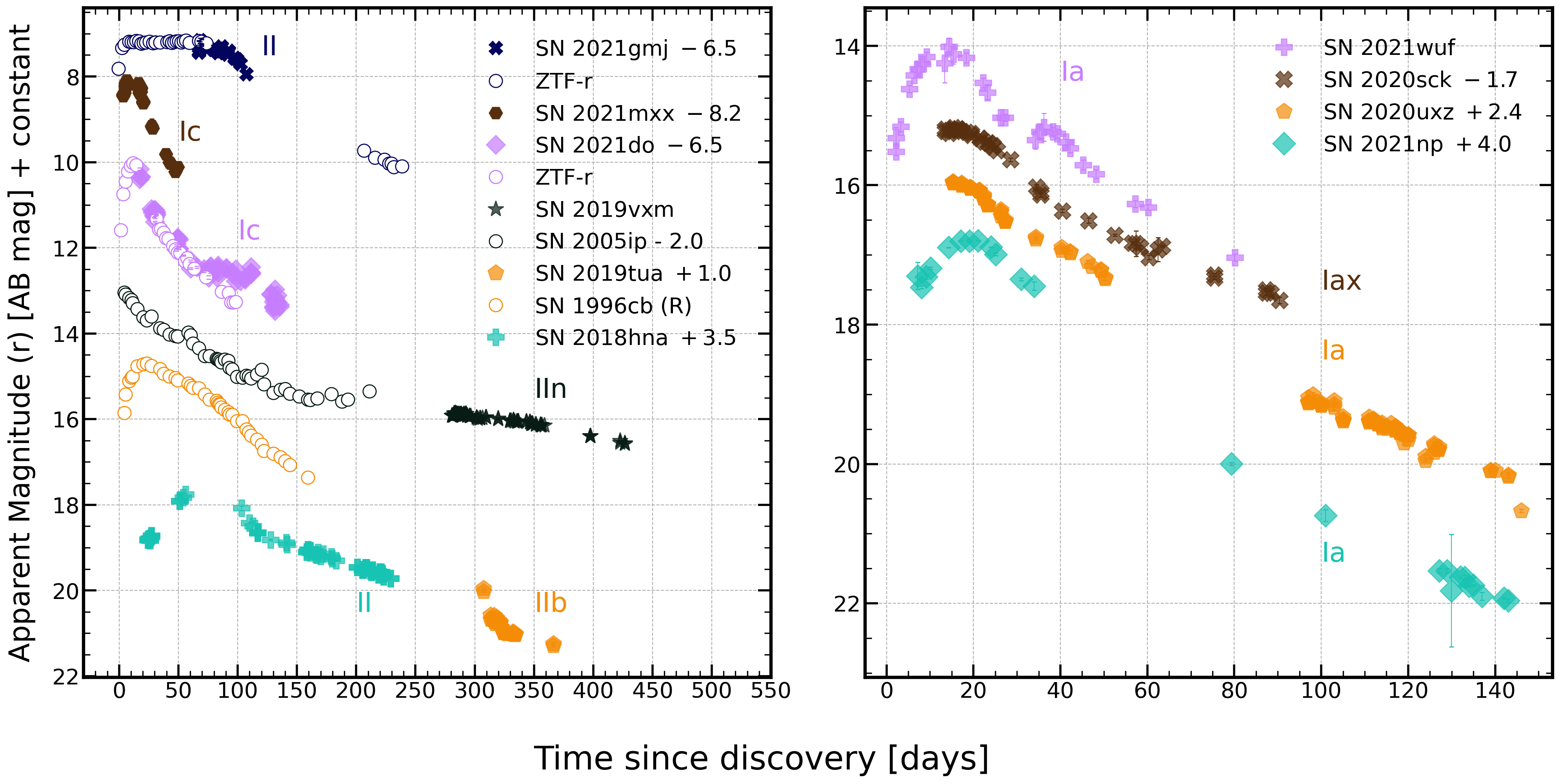}
\caption{$r$-band light curves of a sample of supernovae observed with the GIT. The left panel shows the core-collapse objects, while the right panel shows the thermonuclear SNe. For SN~2019vxm and SN 2019tua, GIT has missed the peak, and we show the early light curves of SN~1996cb \citep{1999AJ....117..736Q} and SN~2005ip \citep{2012ApJ...756..173S} to understand their possible type. GIT has complemented Zwicky Transient Facility (ZTF) surveys in some phases. For some SNe like SN~2021do and SN~2021gmj, we show the light curves in the $ZTF-r$ band \citep{2019PASP..131a8002B}.}
\label{fig:supernove}
\end{figure*}

Most of the GRB afterglows ($>75\%$) were followed up within the first night from the trigger with a median response time of 0.55~days (including waiting for sunset). Our fastest turnaround time was 17~mins for GRB~210722A. Further, the independent discovery of the afterglow within the first hour of image acquisition for the GRB~200524A was one of the key highlights of the GRB follow-up effort. We did detailed follow-ups and analysis for a few other GRBs, including GRB~190114C, GRB~190530A, and GRB 210204A. The popular GRB~190114C detected at sub-TeV energies by MAGIC Telescope~\citep{2019GCN.23701....1M}, was extensively follow-up with GIT and results were published in \citet{2021MNRAS.504.5685M}. Results from the analysis of GRB~190530A were recently published in \citet{2022MNRAS.tmp...51G} where we probed into the emission mechanisms of this GRB. Another interesting GRB~210204A which showed deviations from typical synchrotron-driven afterglow emission at late times was followed up by GIT in coordination with other Indian facilities like HCT, DOT, and DFOT. A detailed study of this GRB was published in \citep{GRB210204A-harsh-paper}.

\subsection{Supernovae}

GIT has followed up over 30 SNe of all types based on their visibility. GIT can be triggered immediately after discovery and has the capability to follow up to very late phases ( $>$ 300 days). Very well-sampled light curves have been obtained for key targets. In Figure~\ref{fig:supernove} we show $r'$-band light curves of a sample of core-collapse and thermonuclear SNe.
GIT has often followed up peculiar events, which can yield high scientific returns. For instance, SN~2018hna --- an SN~1987A like explosion with shock breakout and of blue supergiant progenitor --- has been studied in detail in \cite{2019ApJ...882L..15S} (Figure~\ref{fig:supernove}, left). A type Iax event, SN~2020sck, has been studied in \cite{2022ApJ...925..217D}. One-dimensional radiative transfer modeling of SN~2020sck (Figure~\ref{fig:supernove}, right) has shown the explosion to be a pure deflagration in a carbon-oxygen white dwarf. 

In the left panel of Figure~\ref{fig:supernove}, we show GIT data with solid symbols, complemented with publicly available data in open symbols.
For the type-II event SN~2021gmj, GIT data covered a crucial decline phase. The Ic event SN~2021mxx was covered through its peak, while SN~2021do was observed from the peak onwards.
SN~2019vxm and SN~2019tua have been followed up mostly in the late phases. 
To understand the types of these events, we compare the early phase light curve of SN~2019tua with a type IIb event SN~1996cb \citep{1999AJ....117..736Q}. 
Similarly, for SN~2019vxm, we compare with type IIn event SN~2005ip \citep{2012ApJ...756..173S}. 
GIT has complemented surveys like the Zwicky Transient Facility \citep[ZTF;][]{2019PASP..131a8002B} in some phases of evolution of a few objects and can be instrumental in following up in crucial phases of transient evolution.
In the right panel of Figure~\ref{fig:supernove}, we show GIT r-band lightcurves for four thermonuclear supernovae. A paper combining SN~2021wuf and SN~2019np is under preparation.

\subsection{Novae}

GIT has observed several novae, both galactic and extra-galactic. We have monitored 
the outbursts of the remarkable recurrent nova in M31 --- M31N2008-12a --- in 2018 \citep{2018ATel12203....1S}, 2020, and 2021 \citep{2021ATel15038....1B}. We have monitored the 2021 outburst of the Galactic recurrent nova RS Ophiuchi, as well as  
several classical novae, including V1674 Her, AT 2021ypn, and AT 2022uz (or M31N2022-01a). Monitoring novae in M31 has even led to a serendipitous observation of the classical nova, AT2022cpe, as early as an hour from the time of discovery \citep{2022ATel15250....1B}.

\subsection{Electromagnetic Counterparts of Gravitational Wave events (EMGW)}
GIT extensively followed up EMGW events during the third observation run (O3) Ligo-Virgo Collaboration (LVC), which began in April 2019 \citep{2021PhRvX..11b1053A}. The first phase of O3 (aka O3a), which lasted until the end of September, resulted in the detection of 6 Binary Neutron star (BNS) and 9 Black Hole - Neutron star merger candidates (BHNS)\footnote{\url{https://gracedb.ligo.org/latest/}}~\citep{2021PhRvX..11b1053A}. Among these, six events: GW190425/S190425z (second BNS merger event), GW190426$\_$152155/S190426c, GW190814/S190814bv, S190901ap, S190910d, S190910h were followed by GIT. For all but the S190426c event, GIT primarily followed up interesting candidates at the early times of their evolution. The photometric follow-up played an important role in eliminating the non-interesting candidates in search of counterparts of GW events \citep{2020ApJ...905..145K}. Preliminary results of the follow-up were published in the form of GCN circulars~\citep{2019GCN.25632....1K, 2019GCN.24351....1K, 2019GCN.24316....1W, 2019GCN.24304....1W, 2019GCN.24258....1B, 2019GCN.24201....1B}, and detailed analyses were undertaken in collaboration with GROWTH partners \citep{2020ApJ...905..145K, 2019ApJ...885L..19C}. For the `S190426c' event, we worked in close collaboration with Zwicky Transient Facility~\citep[ZTF]{2014htu..conf...27B} and Dark Energy Camera~\citep[DECam]{2015AJ....150..150F} on Blanco Telescope to tile up a significant sky area (22.1~$\rm{deg}^2$ area containing 17.5\% localisation probability) multiple times during the span of $\sim 10$~days. No viable candidate counterpart was detected in GIT follow-up observations. However, the upper limits obtained from follow-up observations helped put stringent constraints on ejecta masses from putative KNe (Kumar et al., in prep).

\begin{figure*}
\centering
\includegraphics[width =0.7\linewidth]{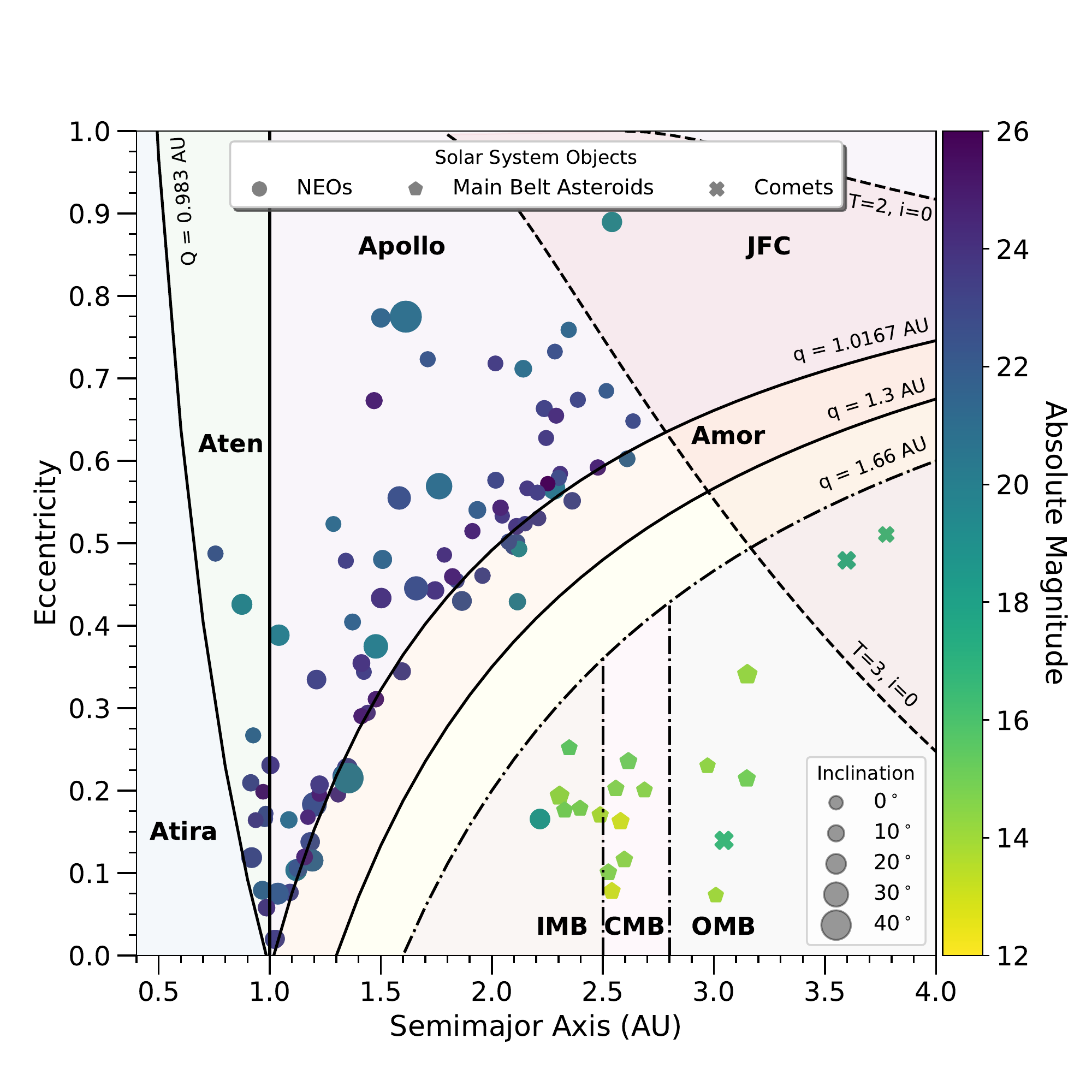}
\caption{An ($a$, $e$, $i$) orbital distribution of all solar system objects observed with GIT. The region between perihelion $q = 1.3$~AU and aphelion $Q = 0.983$~AU demarcates the NEO regime. The orbital parameters of periodic comets with $a > 4$~AU and non-periodic comets are omitted here.}
\label{fig:orbital_distribution}
\end{figure*}

\subsection{Solar System Objects}

GIT has an active follow-up observation program for Near-Earth Objects to refine their orbital parameters. To date, GIT has observed over one hundred NEOs with absolute magnitude up to $H \leq 27.7$\footnote{Full list at \url{https://sites.google.com/view/growthindia/results/asteroids}}. Due to the high proper motion ($3.6 - 120$\arcsec/min), all the NEOs were observed with non-sidereal tracking.

GIT has been used to study seven periodic and sixteen non-periodic comets. We have reported the cometary activity of four non-periodic comets: C/2021 A6 ~\citep{MPEC2021-A210}, C/2021 C1~\citep{MPEC2021-D102}, C/2021 C3~\citep{MPEC2021-D112}, Comet C/2021 E3~\citep{MPEC2021-J71}. In addition to the expected variation in brightness with the geocentric and heliocentric distances, a few comets showed  outbursts i.e., short-term flare-up in luminosity \citep{1975QJRAS..16..410H}. We undertook photometric follow-up of several cometary outbursts discovered by ZTF for confirmation of the event and further follow-up, including Comet 29P/Schwassmann-Wachmann 1 \citep{2021ATel14984....1S, 2021ATel14898....1S, 2021ATel14543....1K}, Comet C/2020 R4 \citep{2021ATel14618....1K}, Comet 22P/Kopff \citep{2021ATel14565....1K, 2021ATel14628....1K}, 67P/Churyumov–Gerasimenko \citep{2021ATel15053....1K, 2021RNAAS...5..277S}.

 (6478) Gault, an active asteroid (originally designated as 1998 JC$_1$) is dynamically a main-belt asteroid with Tisserand parameter with respect to Jupiter $(T_J) = 3.46$ \citep{TisserandParameterforGault}. In June 2020, we participated in coordinated observations with the GROWTH network to constrain its rotation period and understand its activity mechanism. With the help of coordinated observations, we measured a rotation period of $\sim 2.5$~hour~\citep{purdum2021timeseries}, which is the critical rotation period of a body of the size of Gault. It was concluded that the activity of Gault was due to surface mass shedding from its fast rotation spun up by the Yarkovsky-O’Keefe-Radzievskii-Paddack (YORP) effect \citep{Bottke_2006, Kleyna_2019}.

\subsection{Fast transients}
A key component of GIT science involves the follow-up of rare, fast transients. One such event was AT2020xnd/ZTF20acigmel (the ``Camel"), which evolved very fast, similar to the AT2018cow (the ``cow") event~\citep{2019MNRAS.484.1031P}. The candidate maintained a high photospheric temperature even at a later stage of its evolution and with the absence of a second radioactive peak~\citep{2021MNRAS.508.5138P}. We performed late-time deep follow-up in coordination with ZTF and Liverpool optical telescope (LT). The detailed analysis of this event was published in~\citep{2021MNRAS.508.5138P}. GIT followed up another interesting event AT2019pim/ZTF19abvizsw~\citep{2020ApJ...905...98H} during Oct 2019. The candidate showed properties similar to an afterglow of a GRB. However, no GRB was detected for this event by high-energy monitors. The study of this orphan afterglow candidate is in progress and will soon be published in Perley et al. (in prep).

\section{Summary}
GROWTH-India Telescope is India's first robotic telescope, located in Hanle. The telescope has been dedicated to time-domain astronomy since its commissioning in the summer of 2018. After starting its operation in manual mode for a few months, the system was upgraded in a phased manner to semi-automated mode and finally to an automated mode in September 2021. These upgrades significantly increased the telescope's ``on-sky efficiency" to $> 85\%$. The automated bots handle monitoring of the system in real-time and inform the core team of live-time observing updates and errors in case of automatic debug failure. 

GIT has contributed to the time domain astronomy in the last few years by follow-up of EMGW events, GRBs, SNe, Novae, and NEOs. More than 30 core-collapse and thermonuclear SNe have been followed up in the last three years of observations. We have detected 20 GRB optical afterglows and provided useful constraints on the optical brightness of $> 50$ GRB afterglows. We have performed an extensive follow-up of GW events from the O3 run of the advanced gravitational wave detector network, where we chose interesting candidates for follow-up and helped rule them out based on their evolution. During this run, we undertook an extensive transient search operation for S190426c that covered $> 17.5\%$ of the total localisation probability, which eventually constrained the mass ejection models for KNe. Follow-up of NEOs is a regular objective for GIT, where we have observed more than a hundred near-earth objects and discovered/co-discovered four comet outbursts. 

In addition to science, GIT has also been used for education --- an important goal of the GROWTH network. The first international GROWTH school was conducted in India, including a live remote observing session at GIT\footnote{\url{https://www.growth.caltech.edu/growth-winter-school-2018.html}}. The telescope contributes greatly to student training via courses, research projects, as well as undergraduate and graduate theses. 

GIT's automated operations and remarkable observing efficiency give it a unique place in Indian astronomy. In the near future, we are exploring various upgrades to operations, data processing, and hardware to further enhance the observatory's performance. Equipped with these, GIT will continue to play an important role in time-domain astrophysics.

\software{Astropy \citep{astropy:2013, astropy:2018}, Numpy \citep{numpy}, Matplotlib \citep{matplotlib}, Astro-SCRAPPY~\citep{2019ascl.soft07032M}, Flask \citep{grinberg2018flask}, SExtractor \citep{Bertin_1996}, PSFEx \citep{bertin11}, SWarp \citep{SWARP_Bertin}, pyzogy \citep{pyzogy_2017}, astrometry.net \citep{Lang_2010}
}

\section*{Acknowledgements}
The GROWTH India Telescope (GIT) is a 70-cm telescope with a 0.7-degree field of view, set up by the Indian Institute of Astrophysics and the Indian Institute of Technology Bombay with support from the Indo-US Science and Technology Forum (IUSSTF) and the Science and Engineering Research Board (SERB) of the Department of Science and Technology (DST), Government of India. It is located at the Indian Astronomical Observatory (Hanle), operated by the Indian Institute of Astrophysics (IIA). We acknowledge funding by the IITB alumni batch of 1994, which partially supports operations of the telescope. Telescope technical details are available at \url{https://sites.google.com/view/growthindia/}.

Harsh Kumar thanks the LSSTC Data Science Fellowship Program, which is funded by LSSTC, NSF Cybertraining Grant \#1829740, the Brinson Foundation, and the Moore Foundation; his participation in the program has benefited this work.

This research has made use of the NASA/IPAC Extragalactic Database (NED), which is funded by the National Aeronautics and Space Administration and operated by the California Institute of Technology. 
This research has made use of NASA's Astrophysics Data System. 
This research has made use of data and/or services provided by the International Astronomical Union's Minor Planet Center. 
This research has made use of the VizieR catalogue access tool, CDS, Strasbourg, France (DOI : 10.26093/cds/vizier). The original description of the VizieR service was published in 2000, A\&AS 143, 23.
This research made use of Astropy,\footnote{http://www.astropy.org} a community-developed core Python package for Astronomy \citep{astropy:2013, astropy:2018}. 

\bibliography{reference}
\bibliographystyle{aasjournal}
\end{document}